\documentclass[doublecol]{epl2}
\usepackage{epsf}
\usepackage{epsfig}
\usepackage{color}
\usepackage{psfrag}
\usepackage{verbatim}
\usepackage{amsmath}
\usepackage{amsfonts}
\usepackage{appendix}
\usepackage{graphicx}
\usepackage{amssymb}
\usepackage{dsfont}

\title{Effective potential and polaronic mass shift in a trapped dynamical impurity--Luttinger liquid system}

\author{J. Bonart\inst{1} \and L. F. Cugliandolo\inst{1}}
\institute{
  \inst{1} Laboratoire de Physique Th\'eorique et Hautes Energies,
  Universit\'e Pierre et Marie Curie -- Paris VI - 4 Place Jussieu, 75252 Paris Cedex 05, France
}


\abstract{
Recent experiments with cold atoms on the impurity motion in one-dimensional liquids of interacting bosons have revealed an interesting interplay between the polaronic 
impurity mass shift and the renormalization of the optical potential. We show that the influence of the external trap on the Bose gas leads to a steeper effective potential for the impurity. We propose a framework in which this potential renormalization and the mass shift can be quantitatively understood by combining a semi-classical theory 
of density wave excitations in the Luttinger liquid with the non equilibrium formalism of a quantum Brownian particle. The obtained theoretical results reproduce well recent experimental data.
}

\begin{document}

\maketitle

Diffusion in low dimensional quantum liquids has been a major field of research in the last decade \cite{Fisher95,Damski06,Hakim97,FukuharaGiamarchi12,Kamenev12,Pita04,Caux09,Caux11,Giamarchi03,Giamarchi11,Zvonarev12}. 
In one-dimensional (1D) liquids of interacting bosons a moving impurity is subject to a drag force \cite{Pita04} and dissipates energy for all velocities even at zero 
temperature~\cite{Caux11}.  The experimental design of artificial 1D impurity--quantum liquid systems has now become possible by confining cold atoms in optical nanotubes. 
Using these techniques, the diffusion of impurity atoms in contact with a Luttinger liquid (LL) with tunable impurity-LL interaction
was recently studied~\cite{Giamarchi11}. Due to the external trapping potential the minority atoms undergo damped oscillations which directly confirm 
that dissipation takes place in this system. 

In~Ref.~\cite{Johnson12} the authors attempted at a theoretical description of this experiment with the Gross-Pitaevskii formalism. In this Letter we pursue an alternative approach by considering 
the impurity atom as a quantum Brownian particle. The quantum liquid then plays the role of an exotic quantum bath and it can be dealt with using Luttinger theory. As 
already pointed out in \cite{Giamarchi11,BonartCugliandolo12,Johnson12} the impurity atom acquires an effective mass due to its interaction with the LL. 
Moreover, the external trapping potential leads to an inhomogeneous density profile of the LL with non-trivial effects on the impurity motion. In this Letter we 
combine several independent ideas that will allow us to: (a) estimate the mass shift of the impurity, (b) evaluate the effect of the non-homogeneous density profile of the LL
as, to a first approximation, the renormalization of the confining potential, 
(c) use  the non-equilibrium formalism of quantum Brownian motion developed in~\cite{BonartCugliandolo12}
to  reproduce the data in~\cite{Giamarchi11} \emph{quantitatively}.
      
The free Hamiltonian of an impurity with mass $M_I$, in an optical trap modeled by a harmonic potential 
with spring constant $\kappa$, reads
\begin{equation}
\hat{\mathcal H}_I = \frac{\hat p^2}{2 M_I} + \frac{\kappa}{2} \hat q^2 
\ ,
\label{eq:H_I}
\end{equation} 
with $\hat p$ and $\hat q$ the momentum  and position operators.  

The impurity interacts with an LL which is confined by the same potential \cite{Giamarchi11}. For the sake 
of simplicity we will incorporate the effects of the trap on the LL later. The low energy excitations of an unconfined 1D quantum liquid are described 
by the Tomonaga-Luttinger Hamiltonian
\begin{align}\label{eq:hamX}
\hat{\mathcal H}_L &= \frac{\hbar}{2\pi}\int {\rm d} x \left[\frac{uK}{\hbar^2}(\pi\hat\Pi(x))^2+
\frac{u}{K}(\nabla\hat\phi(x))^2\right] 
\nonumber\\
&= \sum_{k\ne 0}\hbar u|k|\hat b_k^\dagger\hat b_k
\ , 
\end{align} 
with the two canonically conjugate bosonic fields $\hat\Pi(x)$ and $\hat\phi(x)$ \cite{Giamarchi03,Cazalilla11}. $\hat\phi$ is related to the LL particle density through 
$\hat\rho(x) = \rho_0 - (1/\pi)\hat\phi'(x)$. In the second quantization language the Hamiltonian can be equally expressed in terms of bosonic  operators 
$\hat b_k^\dagger$ and $\hat b_k$ which we define below. 
The dimensionless coefficient $K$ and the sound velocity $u$ totally characterize the low energy properties of such a 1D system. For translationally invariant cases 
they only depend on the Lieb-Liniger parameter
$
  \gamma = M_Lw_L/\hbar\rho_0
$, 
with $M_L$ the mass of the bosons, $\hbar w_L$ the strength of the interaction and $\rho_0$ the density of the LL \cite{Cazalilla11}. 

We model the impurity-LL interaction through
\begin{equation}\label{eq:hamInt}
\hat{\mathcal H}_{IL} = \int{\rm d} x {\rm d} y\ U(x-y)\hat\rho(y)\delta(x-\hat q)
\ ,
\end{equation}
with $U(x)$ the interaction potential, $\hat\rho(x)$ the LL density and $\hat q$ the impurity position operator. 
In Fourier space we define $\hat\phi(x) = L^{-1/2}\sum_k e^{-ikx}\hat\phi_k$ with $k = 2\pi n/L$ and $n\in {\mathbb Z}$. The full 
Hamiltonian is then 
$\hat{\mathcal H} = \hat{\mathcal H}_L + \hat{\mathcal H}_{IL} + \hat{\mathcal H}_{I}$, with $\hat{\mathcal H}_I$ defined in 
eq.~(\ref{eq:H_I}) and 
\begin{eqnarray}
\label{eq:H_L}
\hat{\mathcal H}_L & =  &
\frac{u}{2}\sum_k\left[\hat\Pi_k\hat\Pi_{-k}+k^2 \hat\phi_k\hat\phi_{-k}\right] 
\ , 
\\
\label{eq:H_IL}
\hat{\mathcal H}_{IL} & =  & 
\sqrt{\frac{K}{\pi\hbar}}\sum_k ik U_k \hat\phi_k e^{-ik\hat q} 
\ .
\end{eqnarray}
Note that we rescaled the fields according to $\hat\phi_k \mapsto \sqrt{(\pi K/\hbar)} \hat\phi_k$ and $\hat\Pi_k \mapsto \sqrt{(\hbar/\pi K)}\hat\Pi_k$. In terms of $\hat b^\dagger_k$, $\hat b_k$ 
the (rescaled) field reads $\hat \phi_k = \sqrt{\hbar/2|k|}(b_{-k}^\dagger + b_k)$.
We choose $U_k = \hbar w / \sqrt{L} \ e^{-u|k|/2\omega_c}$ with some cutoff wave vector $\omega_c/u$ depending on the microscopic properties of the coupling. Equation~(\ref{eq:H_IL}) considers only the so-called forward impurity-LL scattering. The backward scattering potential is not relevant in our case since we consider light impurities~\cite{Caux11,Kamenev12}. 

In \cite{Giamarchi11} the impurities are initially localized at the center of the potential tubes with a laser blade that creates another harmonic potential 
well with spring constant $\kappa_0>\kappa$. After their subsequent release they undergo stochastic 
dynamics that resemble the ones of a damped harmonic oscillator. Catani {\it et al.} measured the equal-time correlation function $\mathcal C(t,t)$ and they drew the following conclusions:

(I) The oscillation frequency $\Omega_I$ is virtually not affected by the value of the impurity-LL interaction $\hbar w$. 
 
(II) Equations~(\ref{eq:H_I})-(\ref{eq:H_IL})  resemble the well-known Fr\"ohlich polaron Hamiltonian that should result in the impurity mass renormalization, $M_I \mapsto M_I^*$, as a function of the 
interaction $\hbar w$. 
Point (I) then indicates that in parallel to the mass renormalization the potential spring constant  should be renormalized as well, $\kappa\to \kappa^*$, 
in such a way that $\Omega_I^* \equiv \sqrt{\kappa^*/M_I^*}$  remained equal to $\Omega_I$. 

\begin{figure}
  \begin{center}
  \includegraphics[width=0.42\textwidth]{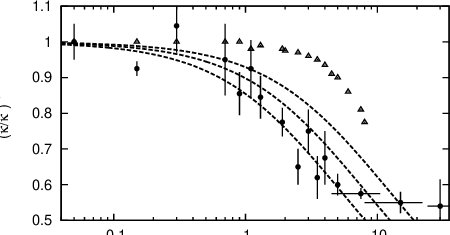}
  \end{center}
  \caption{$\sqrt{\kappa/\kappa^*}$ versus the impurity-LL coupling $w/w_L$. Lines obtained from (\ref{eq:masseff}) with (from left to right) $\gamma = 0.25$, $\gamma = 0.35$ and $\gamma=0.5$. Experimental data points are taken from \cite{Giamarchi11}. Triangular 
  points: Result from Feynman's variational theory \cite{Giamarchi11}.}
  \label{fig1}
\end{figure}

(III) The initial kinetic energy of the impurity can be estimated from the high temperature equipartition theorem to be $\sim 1/\beta$ (note that $\hbar\beta\sqrt{\kappa_0/M_I} \approx 0.1$ in \cite{Giamarchi11}) by assuming that the impurity has equilibrated with the LL before its release. The amplitude after one oscillation $q_a$ should therefore scale as $\sim 1/\sqrt{\kappa^*}$ when neglecting dissipation such that $\kappa^* q_a^2 \sim 1/\beta$ due to energy conservation. Furthermore, $\sqrt{\kappa/\kappa^*} \sim \sqrt{M_I/M_I^*}$ due to point (I). The increase of $\kappa^* \sim M_I^*$ is clearly observed when $\hbar w$ is ramped up 
(see Fig.~\ref{fig1}). Note that for $w/w_L \gtrsim 5$ the 1D regime is not ensured any longer which explains the ``saturation'' of $\kappa^*$ for large values of $w$ 
(not described by the effective 1D theory).  

(IV) The final (equilibrium) width of the impurity cloud is independent of $\hbar w$. 

We notice that  (IV) is at odds with (II). From the theory of quantum Brownian motion we know that $\lim_{t\to\infty}\mathcal C(t,t) \simeq 1/(\beta\kappa^*)$ for a 
harmonic potential with spring constant $\kappa^*$ \cite{Grscin88}; therefore,  the dependence of $\kappa^*$ on $\hbar\omega$ should entail a dependence
of the cloud width with the same parameter.
Accordingly, a more thorough analysis of the coupled systems is needed to correctly interpret the experimental evidence. In the following we examine the points (I)-(IV)  in detail
and we propose a way out this {\it conundrum}.

\emph{The dynamical mass shift}.
It is well-known that a charged particle acquires an effective mass when it interacts with lattice vibrations through a Coulomb potential. 
Equations~(\ref{eq:H_L})-(\ref{eq:H_IL}) describe such a 
polaron with the only difference that the interaction is not Coulomb-like. In the following we estimate the dynamic polaronic 
mass shift in our problem 
by using the equations of motion (EOM) for  $\hat\phi_k(t)$ and 
$\hat q(t)$ (see \cite{Devreese75} for the use of EOM in this context):
\begin{eqnarray}
\label{eq:phi_k}
&&  \ddot{\hat\phi}_k(t) + u^2k^2\hat\phi_k(t) = iuk\sqrt{\frac{K}{\pi\hbar}} U_k^*e^{ik\hat q(t)}
\; , 
\\
\label{eq:EOMq}
 && M_I \ddot{\hat q}(t) + \kappa \hat q(t) =
 \nonumber\\
 && \qquad\qquad\;\;
  - \sqrt{\frac{K}{\pi\hbar}}\sum_k k^2 U_k e^{-ik\hat q(t)}\hat\phi_k(t) \; .
\end{eqnarray}

Suppose that the impurity is not accelerated during a small time interval, then we can make the Ansatz 
\begin{equation}\label{eq:ansatz}
\hat q(t) = \hat q(0) + \hat v t
\end{equation}
with $\hat v = \hat p(0)/M_I$ and $[\hat q(t),\hat\phi_k(t)] \approx 0$. The solution to eq.~(\ref{eq:phi_k}) 
for small $t$ reads
\begin{equation}\label{eq:wave}
  \hat\phi_k(t) = \hat A_k(t;\hat v) e^{ik\hat v t} + \hat g_k e^{i u k t} + \hat h_k e^{-i u k t}\; ,
\end{equation}
with the coefficients
\begin{eqnarray}
&&  \hat A_k(t; \hat v) = i u k \ U_k^*  \sqrt{\frac{K}{\pi\hbar}}
\ \frac{e^{ik\hat q(0) + i\hbar k^2 t/2M_I}}{u^2 k^2 - \hat v^2 k^2}
\ ,  
\\
&&  \hat g_k = 
  \frac{1}{2}\left[\hat\phi_k(0)+\frac{1}{iuk}\dot{\hat\phi}_k(0) - \hat A_k(0;\hat v) -  \hat A_k(0; \hat v)\frac{\hat v}{u}\right] 
  ,\nonumber 
  \\
&&   \hat h_k = \frac{1}{2}\left[\hat\phi_k(0)-\frac{1}{iuk}\dot{\hat\phi}_k(0) - \hat A_k(0;\hat v) +  \hat A_k(0;\hat v)\frac{\hat v}{u}\right] \; .
 \nonumber
\end{eqnarray}
The first term on the right-hand-side of (\ref{eq:wave}) represents a density cloud that moves together 
with the impurity, while the two last terms 
describe the very wave excitation. For instance, in the limit $\omega_c \to \infty$ we have for a mobile impurity with \emph{constant} velocity 
\begin{equation}
\hat\rho(x,t)\sim\sum_k ik\hat A_k(t;\hat v) e^{ik\hat v t - i k x} \sim \delta(x-\hat q(t))
\; , 
\end{equation} 
meaning that the LL density profile follows the 
impurity, thus creating a \emph{dressed} local impurity. It is important to note that this simple picture has to be altered when 
the impurity is accelerated. We will come back to this point later. 

The combined system of the impurity and the density cloud has an effective mass which exceeds the bare impurity mass. As an illustration, we consider the initial conditions $\hat \phi_k(0) = \hat A_k(0;0)$ and $\dot{\hat\phi}_k(0) = 0$. If the impurity is immobile [i.e. $\hat q(t) = \hat q(0)$] these initial conditions lead to the static solution $\hat\phi_k(t) = \hat A_k(0;0)$ which describes a static density cloud without wave excitations. Suppose now that the impurity is \emph{instantaneously} accelerated to some constant velocity $\hat v$, then we obtain from (\ref{eq:wave}) $\hat\phi_k = \hat A_k(t,\hat v) e^{ik\hat v t} - i\hat A_k(0;0) \frac{\hat v}{u} \sin u k t$. Hence, upon acceleration energy is carried away by a wave excitation, such that the kinetic energy of the impurity is less than the external energy provided. To be more specific, by using (\ref{eq:H_L}) the average energy of such a wave excitation is found to be $E_k = \frac{uk^2}{2} (v/u)^2 A_k(0;0) A_{-k}(0;0)$ with $v^2 = \langle \hat v^2 \rangle$. We define the dynamical effective impurity mass through
\begin{equation}\label{eq:masseff}
  M_I^* = (1 + \mu) M_I \; ,
\end{equation}
with the interaction-dependent correction
$\mu = 2\hbar w^2 K \omega_c/(\pi^2 M_I u^4)$. Then, by instantaneously providing an amount of energy $E$, the impurity acquires after acceleration a (mean) velocity given by $E = M_I^* v^2 / 2$. It is straightforward to generalize this calculation to the case where the impurity has already a velocity $\hat v_0$ before the acceleration: One simply replaces $\mu$ by 
\begin{equation}\label{eq:mu}
\mu(v_0) = \frac{\mu}{(1-v_0^2/u^2)^2} \; , 
\end{equation}
where we used the classical mean value $v_0^2$ instead of $\hat v_0^2$. In the following we consider $M_I^*$ as the true impurity mass. Note, that our definition of the dynamical effective mass differs from the effective mass usually defined via the impurity self-energy diagram. 

\emph{The potential renormalization}.
\label{sec:potential}
Ref.~\cite{Giamarchi11} indicates that the spring constant of the optical trap is renormalized as well. We will show here that the effect of the external potential on the LL indeed 
leads to a renormalization of the potential felt by the impurity. In the same spirit as in the previous paragraph we study the effects of the external potential, which we previously neglected, \emph{via} its action on the density cloud. The force exerted by the external harmonic potential on the density cloud is given by
\begin{align}\label{eq:force}
  \hat F = - \int_{-L/2}^{L/2}{\rm d} x \ \kappa x \hat\rho(x)
  = \kappa \hat q \ \frac{u K}{\pi \hbar(u^2-{\hat v}^2)}\sqrt{L}U_0^*  \; ,
\end{align}   
where we used (\ref{eq:wave}) to express $\hat\rho(t)$. Note that the divergency for $v \to u$ stems from the fact that backscattering cannot be neglected when $v$ approaches the speed of sound. (\ref{eq:force}) is therefore expected to be right only for very large or very small $v$. By considering the combined  impurity and the density cloud system
as one entity, $\hat F$ acts in the end on the impurity itself. Interestingly enough, $\hat F$ changes sign when $\hat v$ exceeds the sound speed $u$ such that the subsonic and supersonic regimes are \emph{qualitatively} different. In \cite{Giamarchi11} the impurity moves with supersonic speed ($\sqrt{\langle \hat v^2 \rangle} \approx 8.5 \rm mm/s$ while $u \approx 3 \rm mm/s$ typically) so that $\hat F$ leads to an \emph{increase} of the effective external potential. In the following we approximate $\hat v^2$ by its mean value $v^2 \equiv \langle \hat v^2 \rangle$. 

Intuitively, the potential renormalization can be easily understood. When the impurity creates a density exciton it has to push the LL atoms up the optical potential to be able to create the density cloud. 
Therefore it loses more energy than what the density wave would cost. The inverse is true as well. By absorbing an exciton the impurity gains more energy than 
the exciton provides since potential energy is freed during the absorption process.
%
\begin{figure}
  \begin{center}
    \begin{tabular}{c}
      \includegraphics[width=0.46\textwidth]{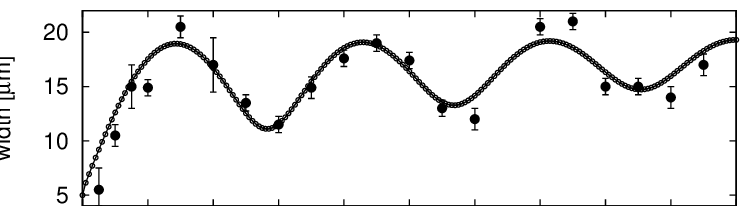} \\
      \includegraphics[width=0.46\textwidth]{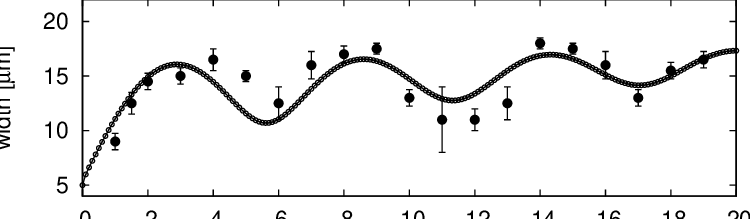}
    \end{tabular}
  \end{center}
  \caption{$\sqrt{C(t,t)}$ for $\Omega_I/\omega_c = 2.5$, $\gamma = 0.45$ and $w/w_L = 1(4)$ in the upper (lower) image. Points: Experimental data from \cite{Giamarchi11}. Lines: Solution to (\ref{eq:Cfinal}) with $\Omega_I^* = \Omega_I$.}
  \label{fig2}
\end{figure}
%
Equation~(\ref{eq:force}) leads to the effective spring constant 
\begin{equation}\label{eq:kappaeff}
  \kappa^* = (1+\tilde\mu(v))\kappa \; ,
\end{equation}
where 
$
\tilde\mu(v) = \frac{K w}{\pi}u/(v^2-u^2) 
$.
In conjunction with eq.~(\ref{eq:mu}) we thus obtain for the effective potential frequency 
\begin{equation}\label{eq:freqeff}
  (\Omega_I^*)^2 = \frac{1+\tilde\mu(v)}{1+\mu(v)}\Omega_I^2 \; .
\end{equation}
In Fig.~\ref{fig1} we compare the prediction in (\ref{eq:kappaeff}) to the experimental data \cite{Giamarchi11}. 
The best curves are obtained for $\gamma \approx 0.25-0.35$. Note that all the constants are determined by the experimental setup. However, since it is difficult to define $\gamma$ for a 
non-homogeneous density we plotted results for $\gamma = 0.5$ and $\gamma = 0.35$ and $\gamma = 0.25$ for illustration. If the non-homogeneous density profile is approximated by an homogeneous one the parameters used in the experimental setup in \cite{Giamarchi11} lead to $\gamma \approx 0.45$. 

\emph{Long time dynamics}:
In the previous two paragraphs we studied the potential renormalization and the impurity mass shift by assuming that the impurity velocity 
was constant. Only then is the impurity cloud perfectly localized around the impurity position. It is clear that such an approximation can only 
hold for short times in the system we consider. For instance, while (\ref{eq:kappaeff}) can still be considered as a realistic approximation up 
to the first oscillation maximum of the impurity, it certainly fails to describe the correct physics for $t \to \infty$. The general solution to 
(\ref{eq:wave}) is  
\begin{equation}
  \hat\phi_k(t) = \int_0^t{\rm d}s \ \frac{\sin u k (t-s)}{u k} e^{i k \hat q(s)} + {\rm wave}\; {\rm excitations} \; ,
\end{equation}
which leads to a density cloud of the form
$\hat\rho(x,t) \sim \int_0^t{\rm d}s \ \delta[x-\hat q(s) - u(t-s)]-\delta[x-\hat q(s) + u(t-s)]$. In the case of an exponentially damped oscillating impurity this density cloud depends only on past values of $\hat q$ when $t$ is large and hence, for $t\to\infty$, the influence of $\hat q(t)$ on $\hat\rho(x,t)$ becomes negligible. 

To put it in other words, for $t\to\infty$ the density cloud is independent of the impurity such that its dynamics decouple from those of 
$\hat q(t)$: The LL has no dynamical effects on the impurity and one concludes that the LL neither renormalizes the impurity mass nor 
the external potential. 
Accordingly, the final width of the impurity reads
\begin{equation}\label{eq:final}
  \mathcal C(t,t) \simeq \frac{1}{\beta\kappa}\;\;\;\mathrm{for}\;t\to\infty 
\end{equation}
and not $1/\beta\kappa^*$. We have thus found that dynamical quantities depend on the renormalized values $\kappa^*$ and $M_I^*$ while final equilibrium quantities have to be computed with the bare values $\kappa$ and $M_I$. 
We insist on the fact that this behaviour has been observed by \cite{Giamarchi11} where the final impurity position 
width is not renormalized in contrast to the potential renormalization that is observed at short times (see Fig.~\ref{fig1}).

\emph{The impurity influence functional}:
We now use the Keldysh formalism to derive an effective out of equilibrium action for the dynamical impurity position. The action of the free oscillator (described by $\mathcal H_I$ with the parameters $M_I^*$ and $\kappa^*$) 
is complemented by
\begin{align}
& \mathcal S_{inf}[q^+,q^-,q^0] = \sum_k \displaybreak[0]\
\Big\{ 
\\
&
-i\int_0^{\beta\hbar} \!\! {\rm d}\tau \! \int_0^\tau \! {\rm d}\sigma \ \Gamma_k(-i\tau+i\sigma)e^{i k q_0(\tau)-i k q_0(\sigma)} 
\nonumber \displaybreak[0]\\ 
& 
+\int_0^{\beta\hbar} \!\! {\rm d}\tau \! \int_0^t \! {\rm d}s \ \Gamma_k^*(s-i\tau)e^{i k q_0(\tau)}
\!\! \left[
e^{-i k q^+(s)}-e^{-i k q^-(s)}
\right] 
\nonumber \displaybreak[0]\\
& 
+i\int_0^t{\rm d}s\int_0^s \!  {\rm d}u
\left[
e^{i k q^+(s)}-e^{i k q^-(s)}
\right] 
\nonumber \displaybreak[0]\\
&
\;\;\;\;\;\;
\times \left. 
\left[
\Gamma_{-k}(s-u)e^{-i k q^+(u)}-\Gamma_{-k}^*(s-u)e^{-i k q^-(u)}
\right] 
\right\}
\; \nonumber\displaybreak[0]\\
&
+ \frac{\hbar\beta\kappa_0}{4}\left[q_i^2 + {q'_i}^2\right],
\label{eq:action1}
\end{align}
where $q^+(s)$, $q^-(s)$ are the dynamical Keldysh branches with $q^+(0) = q_i$ and $q^-(0) = q'_i$, and $q_0(\tau)$ is the path over the 
initial condition (with imaginary time $\tau$) \cite{BonartCugliandolo12}. 
The last line in the right-hand-side of (\ref{eq:action1}) describes the initial localization due to the laser blade which we 
interpreted as an initial position measurement with width $1/\kappa_0\beta$.
Damping stems from the impurity-bath coupling which induces the kernel~\cite{Grscin88}: 
\begin{equation}
  \Gamma_k(\theta) = \frac{K|k||U_k|^2}{2\pi\hbar}\frac{\cosh[u|k|(\beta\hbar/2-i\theta)]}{\sinh[u|k|\beta\hbar/2]}
  \ , 
\end{equation}
with $\theta = s-i\tau$.  
In order to understand the effects induced by the non linear impurity-LL coupling in (\ref{eq:H_IL}) we expand eq.~(\ref{eq:action1}) to second order in $q$. 
The result can be found in \cite{BonartCugliandolo12} and the correlation function can be calculated: 
\begin{equation}\label{eq:corr}
  \mathcal C(t,t) \simeq \frac{\hbar^2\beta\kappa_0}{4} \ \mathcal R^2(t) - \kappa^*\beta \ \mathcal C^{\rm eq}(t)^2 + 
\frac{1}{\kappa^*\beta} \; ,
\end{equation}
Here, $\mathcal R(t)$ and $\mathcal C^{\rm eq}(t)$ are the response and equilibrium 
correlation functions in the high temperature limit $\hbar\beta\Omega_I \ll 1$ which prevails in the experiment. In the Laplace domain they 
read $\tilde{\mathcal R}(z) = (1/M_I^*) [z^2 + z\omega_c\tilde\alpha(z) + (\Omega_I^*)^2 ]^{-1}$ and $\tilde{\mathcal C}^{\rm eq}(z) = 
(1/\beta z)[1/\kappa^* - \tilde{\mathcal R}(z)]$. Linear response and 
correlator depend only on the ``damping kernel'' 
\begin{equation}\label{eq:dampkernel}
  \alpha(t) = \int_0^\infty{\rm d}\omega\ \frac{\mu M_I}{M_I^*} \left(\frac{\omega}{\omega_c}\right)^2 e^{-\omega/\omega_c} \cos\omega t\; .
\end{equation}

As we pointed out, the final equilibrium value should be rather $1/\kappa\beta$ \cite{BonartCugliandolo12} than $1/\kappa^*\beta$ that would follow from (\ref{eq:corr}). We conclude that, while the Gaussian approximation of (\ref{eq:action1}) [see \cite{BonartCugliandolo12} for details] yields a realistic description of the impurity dynamics for short times, it cannot deliver the right correlation function for large times where a crossover from the effective constants $\kappa^*$ and $M_I^*$ to bare quantities takes place. Since our approach does not provide us with an explicit expression of $\kappa(t)$ and $\Omega_I(t)$ we directly construct an approximate correlator
\begin{align}\label{eq:Cfinal}
  \mathcal C(t,t) &\approx \frac{\hbar^2\beta\kappa_0}{4} \ \mathcal R^2(t) - \kappa^*\beta \ \mathcal C^{\rm eq}(t)^2 + \frac{1}{\kappa^*\beta} \nonumber \\
  &+\left(1-e^{-\Gamma\Omega_I t}\right)\left(\frac{1}{\kappa\beta}-\frac{1}{\kappa^*\beta}\right) \; ,
\end{align}
which interpolates between the two asymptotic expressions (\ref{eq:corr}) and (\ref{eq:final}). Here, $\Gamma$ is the effective damping induced by the Luttinger bath. For small to moderate damping it is given by $\Gamma \simeq \frac{\pi}{8}\mu(\Omega_I/\omega_c) e^{-\Omega_I/\omega_c}$ \cite{BonartCugliandolo12}. The explicit form of $\Gamma$ provides us with a physical interpretation of the dissipation process: The forward scattering potential (\ref{eq:H_IL}) does not lead to any friction for an impurity with a constant velocity. However, an external potential changes the impurity dispersion such that a trapped impurity can always emit energy. Indeed, when $\Omega_I \to 0$ the friction $\Gamma$ vanishes. Hence, this \emph{bremsstrahlung} like dissipation is qualitatively different from the backscattering friction discussed in~\cite{Fisher95}.

\emph{Discussion.}
In~\cite{Giamarchi11} $\rm ^{41}K$ atoms play the role of the impurities moving in optical 1D tubes through a Luttinger liquid made of $\rm ^{87}Rb$ atoms. Both the $\rm ^{41}K$ and the 
$\rm ^{87}Rb$ are confined in the same longitudinal optical potential with the (bare) potential frequency $\Omega_I= 550 \rm s^{-1}$ ($390 \rm s^{-1}$) for $\rm ^{41}K$ ($^{87}\rm Rb$). We interpret the initial 
localization (with $\kappa_0 \approx 150 \kappa$) of the impurities as a position measurement. The experimental temperature is such that $\hbar\Omega_I\beta \simeq 10^{-2}$ which ensures the high temperature regime. The mean squared velocity is obtained to be $\sqrt{v^2} \approx 8.5 \ \rm mm/s$ which exceeds the typical sound velocity $u \approx 3 \ \rm mm/s$ so that the impurity moves in the supersonic regime.

As pointed out before, the mass has to be renormalized in such a way that $\Omega_I^*$ remains approximately constant over a wide range of $w/w_L$. This can be achieved by a suitable choice of $\omega_c$, the only free parameter in our theory. For $\gamma = 0.35$ the choice $\omega_c/\Omega_I = 40-50$ leads to a variation of $10\ \%$ for $\Omega_I^*$ in the range $0 < w/w_L < 5$. 
However, since the mass and potential shifts decrease during equilibration the oscillation frequency can slightly change in time. Thus, for large times $\Omega_I^*$ approaches $\Omega_I$ in any case. Hence, in order to experimentally observe the $\Omega_I^*$ predicted by Eq.~(\ref{eq:freqeff}) one cannot average over many periods as was done in \cite{Giamarchi11}. It is therefore not straightforward to make a direct precise quantitative comparaison between Eq.~(\ref{eq:freqeff}) and the experimental findings, although we think that the evidence in \cite{Giamarchi11} clearly indicates that the mass renormalization counteracts the potential shift to a large extent. 
%
%
%

Finally, we compare (\ref{eq:Cfinal}) to experimental data in Fig.~\ref{fig2}. In \cite{Giamarchi11} $\sqrt{\mathcal C(t,t)}$ has an offset of about $5\rm\mu m$ which we add to our theoretical results. Moreover, for small interactions ($w/w_L \lesssim 1$) there is a residual damping in the experiment due to inter-impurity collisions in tubes with several impurity atoms \cite{Giamarchi11} which is of course not covered by our theory. We therefore use  the data from \cite{Giamarchi11} for the damping constant ($\Gamma\approx 0.03$ for $w/w_L = 1$) in $\alpha(t)$. As pointed out before $\Omega_I^*$ can slightly vary during the equilibration process. However, this effect is not expected to be observable within the experimental error bars and therefore we approximate $\Omega_I^*$ by $\Omega_I$ for all times. The match between the experimental data and our theoretic curves is quite impressive.

In a recent paper \cite{BonartCugliandolo12} we argued that non-trivial effects on $\Omega_I$, produced by the super-Ohmic spectral density in the damping kernel (\ref{eq:dampkernel}), could be observed for $\omega_c \approx 0.3\Omega_I$. Now, after having gained a deeper insight into the impurity dynamics we think that $\omega_c$ should be much larger such that the effects of the polaronic mass shift and the potential renormalization (which were previously neglected in \cite{BonartCugliandolo12}) largely dominate the influence of the non-Ohmic spectral density. 

In summary we gained a thorough theoretical understanding of the experimental data in \cite{Giamarchi11}. We calculated the effective 
potential spring constant with an EOM approach, which is expected to be correct for short times, and we obtained a result without any 
undetermined parameter [see Fig.~\ref{fig1}]. We argued that due to memory effects neither a potential nor a mass renormalization can take 
place in the long time limit. One question not resolved yet concerns the precise mechanism that links the mass and potential shifts which we hope will be revealed by future experiments. Finally, using the analytic results in~\cite{BonartCugliandolo12}
for the Brownian motion of a particle coupled to an exotic environment, after an initial position measurement, and with a 
phenomenological correction to the asymptotic limit, we described the experimental data for ${\cal C}(t,t)$ very accurately.

\acknowledgements
 We thank T. Giamarchi and C. Castelnovo for very helpful discussions. This work was financially supported by ANR-BLAN-0346 (FAMOUS).


\bibliographystyle{phjcp}
\bibliography{artbib1.bib}

\end{document}